\documentclass[conference]{IEEEtran}

\IEEEoverridecommandlockouts

\usepackage{cite}
\usepackage{amsmath,amssymb,amsfonts}
\usepackage{graphicx}
\usepackage{textcomp}
\usepackage{xcolor}
\usepackage{multirow}
\usepackage{url}

\usepackage{subcaption}
\usepackage{rotating}

\usepackage{tikz}
\usetikzlibrary{positioning,arrows.meta}
\usepackage[caption=false,font=footnotesize]{subfig}

\title{Temporal Property-driven Design Space Exploration with Reinforcement Learning for Cyber-Physical Systems}

\author{
\IEEEauthorblockN{Tagir Fabarisov}
\IEEEauthorblockA{
\textit{SnT, University of Luxembourg}\\
tagir.fabarisov@uni.lu
}
\and
\IEEEauthorblockN{Maxime Cordy}
\IEEEauthorblockA{
\textit{SnT, University of Luxembourg}\\
maxime.cordy@uni.lu
}
}

\begin{document}

\maketitle

\begingroup
\renewcommand{\thefootnote}{}
\footnotetext{\footnotesize
\textit{Accepted manuscript for IECON 2026---52nd Annual Conference of the IEEE Industrial Electronics Society, Doha, Qatar, 18--21 October 2026.}

\copyright~2026 IEEE. Personal use of this material is permitted.
Permission from IEEE must be obtained for all other uses, in any current or
future media, including reprinting/republishing this material for advertising
or promotional purposes, creating new collective works, for resale or
redistribution to servers or lists, or reuse of any copyrighted component of
this work in other works.
}
\endgroup

\begin{abstract}
Design-space exploration of configurable Cyber-Physical Systems (CPS) requires executable evaluation when design choices affect timing, fault propagation, recovery behavior, and temporal-property satisfaction. Repeated stochastic executions make exhaustive exploration impractical for large design spaces. This paper presents a temporal-property-driven CPS design workflow using Reinforcement Learning (RL). At design time, the RL agent selects subsystem alternatives to assemble a candidate system model. The model is then evaluated through simulation, during which online temporal-property monitors observe runtime traces and produce functional-property violation indicators. These indicators are combined with evaluated non-functional terms for budget, recoverability, sustained compliance, and operational use to calculate the reward used for subsequent candidate selection. The workflow is evaluated on a methane-sensitive mine-pump CPS. The corresponding executable case-study model is provided as additional contribution. RL-guided search identifies the highest-reward design observed in the experiments after 26 episodes (corresponds to 130 executable simulations). These designs were reached with fewer simulations than surrogate-guided Bayesian Optimization and population-based Genetic Algorithm baselines under the same executable model and reward formulation. Ablation study results indicate that value-based feedback and reuse of previous simulation traces contribute to this reduction.
\end{abstract}


\section{Introduction}
Configurable Cyber-Physical Systems (CPS) architectures contain design choices related to component selection, subsystem composition, software-to-hardware mappings, and recovery mechanisms~\cite{adil2024ng}. These alternatives affect both static properties, including cost, and execution-dependent properties, including timing, fault propagation, recovery behavior, and satisfaction of temporal requirements~\cite{dafflon2021challenges}. Design-space exploration cannot rely only on static configuration data. A candidate design may satisfy budget or nominal dependability constraints but violate behavioral requirements once stochastic faults, recovery delays, controller decisions, and subsystem interactions are considered during execution~\cite{zhou2025comprehensive}. Runtime traces and temporal-property monitoring make these effects observable during simulation, but exhaustive evaluation becomes impractical when many design choices must be explored~\cite{morozov2026risk,krichen2026formal}.

This paper presents a Reinforcement Learning (RL)-guided methodology for temporal-property-driven CPS design-space exploration. Candidate system models are assembled from subsystem variants and evaluated through simulation. The simulations include stochastic environmental and fault behavior for uncertainty modeling. The RL agent selects candidate designs at design time and does not control the CPS during operation. During each simulation, online temporal-property monitors produce functional-property violation indicators from runtime signals. These indicators and the evaluated non-functional terms are used to calculate the reward. The design vectors, runtime traces, monitor outputs, and rewards from successful and unsuccessful candidates form evaluation evidence. This evidence can support structured interpretation and subsequent candidate selection in repeated design-exploration cycles. The study examines whether previous simulation traces and temporal-property outcomes can be reused through policy learning to reduce the simulation effort required to reach a high-quality design. RL-guided search is compared with Genetic Algorithm (GA)-based and Bayesian Optimization (BO)-based search under the same executable model and reward formulation. GA uses observed rewards for population selection, BO updates a surrogate model from previous evaluations, and RL reuses stored transitions to update its design-selection policy and value estimates.

\noindent\textbf{Contribution.} The contributions are: \textit{(i)} an RL-guided workflow for temporal-property-driven CPS design-space exploration, comprising design-space configuration, behavioral simulation, online temporal-property monitoring, reward calculation, and candidate-design selection; \textit{(ii)} an executable MATLAB/Simulink CPS case study with dependability- and recoverability-related subsystem variants and stochastic environment simulations for uncertainty modeling. The evaluation compares RL-guided search with GA and BO and includes ablation experiments on value-based feedback and reuse of previous simulation traces. RL reaches the highest-reward design observed in the experiments after 26 episodes, corresponding to 130 executable simulations. BO achieves the highest final rewards among the search baselines, whereas GA provides rapid initial progress but more frequently converges to lower-reward regions.

\section{Research gap}
\label{sec:research-gap}
Design-space exploration for configurable CPS must account for interacting choices related to components, sensing units, controller logic, communication links, software-to-hardware mappings, and recovery mechanisms. Properties such as cost or nominal component reliability can be estimated from configuration data~\cite{wijaya2025performance,cederbladh2024early}. Timing, fault propagation, recovery behavior, and temporal requirement satisfaction depend on execution under operating conditions. Configurations with similar static attributes may produce different execution traces and property-satisfaction outcomes~\cite{dafflon2021challenges}. Trace-based temporal-property evaluation addresses this problem by executing candidate designs and monitoring their runtime behavior~\cite{zhou2025comprehensive,krichen2026formal}. However, repeated executions may be required when fault occurrence, recovery timing, or environmental behavior vary across runs~\cite{calvanese2022verification,rozier2007ltl}. Exhaustive evaluation therefore becomes impractical for design spaces with many variation points~\cite{vigano2019smt}. A search mechanism is needed to select the next candidate designs and regions for further refinement~\cite{morozov2026risk,wijaya2025performance}.

Existing methods for reducing the number of evaluated candidates include analytical models, heuristic sampling, metaheuristic search, BO, and learning-based approaches~\cite{taha2025optimizing,naik2023machine,cordy2023towards,deshmukh2017testing}. Metaheuristic methods use simulation outcomes to rank candidate designs by their observed rewards~\cite{banerjee2025challenges}. BO constructs a surrogate model to guide future evaluations~\cite{deshmukh2017testing}. Temporal-property verification and monitoring determine whether runtime behavior satisfies the required properties, but do not define how observed violations and satisfaction outcomes should influence subsequent design decisions~\cite{bartocci2021adaptive}. This paper addresses this gap by coupling temporal-property evaluation with RL-guided design exploration. The proposed approach records evidence from successful and unsuccessful executable evaluations, including simulation traces and temporal-property outcomes. The corresponding state--action--reward experience is reused to guide subsequent design choices instead of evaluating configurations exhaustively or relying only on static design information.

\textbf{Hypothesis.}
RL is evaluated for temporal-property-driven CPS design-space exploration as a mechanism for reusing experience from previous executable evaluations to guide design choices. GA and BO also reuse information from previous evaluations through reward-based ranking and surrogate modelling, respectively, but they do not learn a design-selection policy from accumulated state--action--reward experience. The hypothesis is that this policy-based reuse reduces the number of executable simulations required to reach a high-quality design. Section~\ref{sec:experiments} evaluates this hypothesis through comparisons with GA-based and BO-based search and through ablation studies of RL mechanisms.

\section{Motivating example}
\label{sec:motivating-example}
The motivating example is a mine-shaft pumping CPS based on the mine-pump benchmark scenario commonly used for variability-aware and temporal-logic-based verification of CPS behavior~\cite{classen2012featured,cordy2023towards}. The implemented case study includes stochastic environmental behavior. The system represents a water extraction setup for underground mining environments. It contains a water pump, methane sensing, water-level sensing, and a controller. Its operation combines two objectives: water must be removed from the shaft when the water level requires pumping, and pump operation must be prevented when methane concentration exceeds the safety threshold. The executable mine-pump case-study model is made publicly available as a contribution of this work. \footnote{\label{fn:mathworks-model}The introduced mine-pump case study model is available at: \url{https://www.mathworks.com/matlabcentral/fileexchange/184555-mine-pump-cps-case-study-for-temporal-property-driven-dse}. The model builds on FIBlock, available at: \url{https://www.mathworks.com/matlabcentral/fileexchange/75539-fault-injection-block-fiblock}.} The current implementation is the first version of a case study that will be extended in subsequent work. The case study will serve as a testbed for next-generation CPS with dynamic uncertainty, context shifts, and distributed sensing. Such systems require the more advanced exploration methods discussed in Section~\ref{sec:conclusion}~\cite{morozov2026risk}.

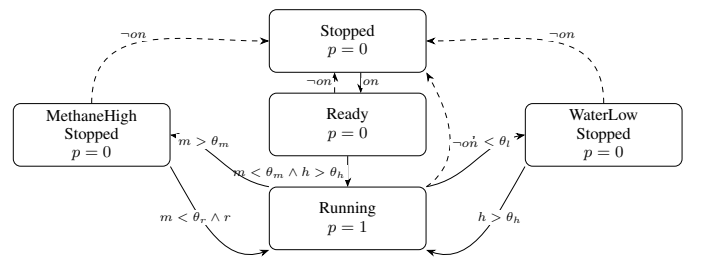
\begin{figure}[b]
\centering
\resizebox{\columnwidth}{!}{%
\begin{tikzpicture}[
    >=Stealth,
    state/.style={
        draw,
        rounded corners,
        align=center,
        font=\small,
        minimum width=30mm,
        minimum height=12mm,
        inner sep=2.5pt
    },
    trans/.style={->, thin},
    offtrans/.style={->, thin, dashed},
    lab/.style={font=\scriptsize, inner sep=1pt, fill=white}
]

\node[state] (stopped) at (0,3.6) {Stopped\\\(p = 0\)};
\node[state] (ready)   at (0,2.0) {Ready\\\(p = 0\)};
\node[state] (running) at (0,0.2) {Running\\\(p = 1\)};

\node[state] (methane)  at (-4.9,1.8) {MethaneHigh\\Stopped\\\(p = 0\)};
\node[state] (waterlow) at (4.9,1.8)  {WaterLow\\Stopped\\\(p = 0\)};

\draw[trans] ([xshift=2.5mm]stopped.south) --
    node[lab,right] {\(on\)}
    ([xshift=2.5mm]ready.north);

\draw[offtrans] ([xshift=-2.5mm]ready.north) --
    node[lab,left] {\(\neg on\)}
    ([xshift=-2.5mm]stopped.south);

\draw[trans] (ready.south) --
    node[lab,left,pos=0.55] {\(m < \theta_m \land h > \theta_h\)}
    (running.north);

\draw[trans] (running.north west) to[out=170,in=350]
    node[lab,pos=0.68,above] {\(m > \theta_m\)}
    (methane.east);

\draw[trans] (methane.south east) to[out=300,in=200]
    node[lab,pos=0.32,below] {\(m < \theta_r \land r\)}
    (running.south west);

\draw[trans] (running.north east) to[out=10,in=190]
    node[lab,pos=0.68,above] {\(h < \theta_l\)}
    (waterlow.west);

\draw[trans] (waterlow.south west) to[out=240,in=340]
    node[lab,pos=0.32,below] {\(h > \theta_h\)}
    (running.south east);

\draw[offtrans] (methane.north)
    to[out=90,in=180,looseness=1.15]
    node[lab,above left] {\(\neg on\)}
    (stopped.west);

\draw[offtrans] (waterlow.north)
    to[out=90,in=0,looseness=1.15]
    node[lab,above right] {\(\neg on\)}
    (stopped.east);

\draw[offtrans] (running.north east)
    to[out=35,in=300,looseness=1.05]
    node[lab,pos=0.42,right] {\(\neg on\)}
    (stopped.south east);

\end{tikzpicture}%
}
\caption{Control logic of the mine-pump controller. Here, \(p\) denotes pump activation, \(on\) system availability, \(m\) methane concentration, \(h\) water level, \(r\) restart permission, and \(\theta_m\), \(\theta_r\), \(\theta_h\), and \(\theta_l\) denote the methane shutdown, methane restart, high-water, and low-water thresholds, respectively.}
\label{fig:simplified_stateflow}
\end{figure}

The normal operating sequence starts in \textit{Stopped} and moves to \textit{Ready} once the system is online. If methane remains below the threshold and the water level exceeds the high-water threshold, the controller enables pumping and enters \textit{Running}. Pumping stops in \textit{WaterLowStopped} when the water level drops below the low-water threshold. If methane exceeds the safety threshold, the controller enters \textit{MethaneHighStopped}; restart is permitted only after the methane level falls below the restart threshold and restart permission is available. The case study is implemented in Simulink/Stateflow. Stateflow defines the controller modes and transitions, while Simulink represents the physical process, signal propagation, and subsystem interactions. Figure~\ref{fig:simplified_stateflow} presents the corresponding state diagram. Fault injection uses our model-based mechanisms~\cite{fabarisov2021model,fabarisov2023remedy,moradi2026optimization}.

\begin{figure*}[!t]
    \centering
    \includegraphics[width=0.98\textwidth]{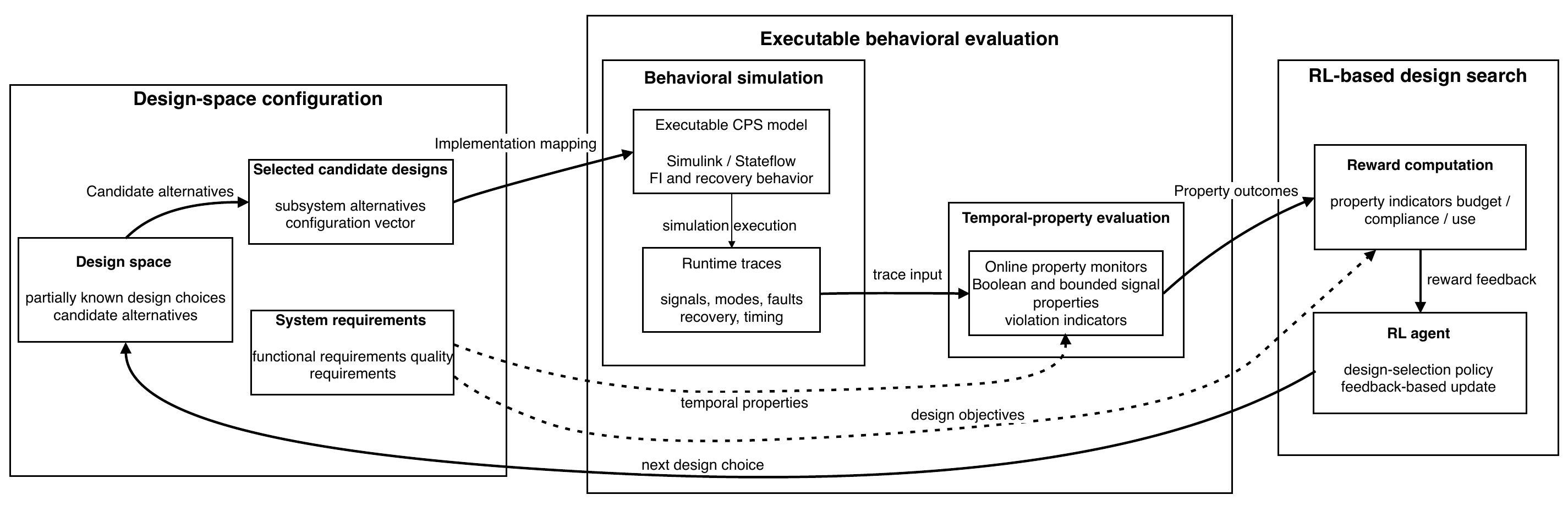}
\caption{Workflow of the proposed RL-guided CPS Design-Space Exploration methodology.}
\label{fig:workflow}
\end{figure*}

\begin{figure*}[!t]
    \centering
    \includegraphics[
        width=0.98\textwidth,
        keepaspectratio,
        trim=0 23mm 0 23mm,
        clip
    ]{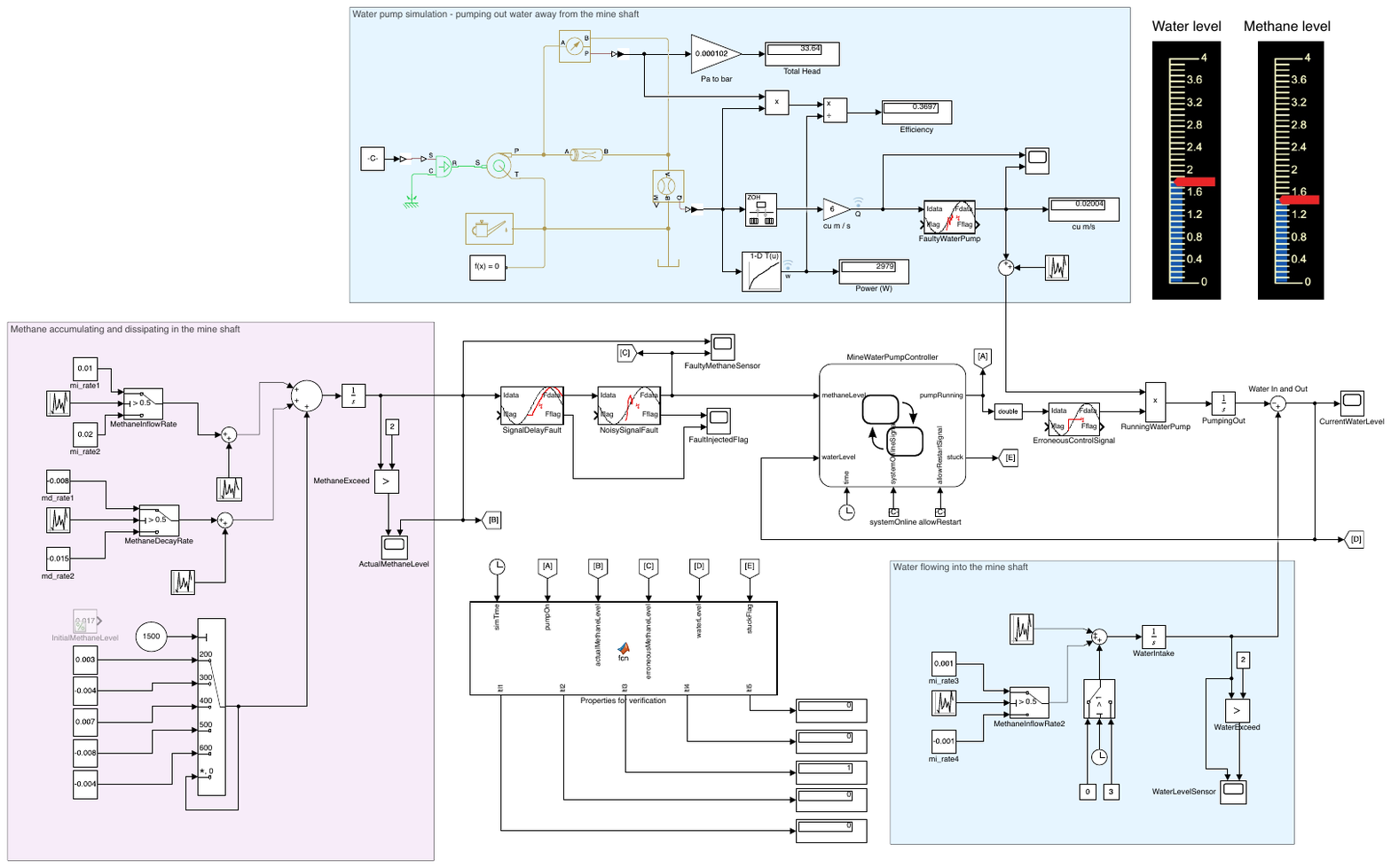}
    \caption{Executable Simulink/Stateflow model of the mine-pump CPS used for temporal-property-driven design-space exploration. The model includes the physical process, sensing and actuation paths, model-based fault-injection mechanisms, Stateflow control logic, temporal-property monitors, and reward-related signals.}
    \label{fig:minepump}
\end{figure*}

\section{Proposed methodology}
\label{sec:methodology}

The proposed methodology addresses CPS design-space exploration as a learning-based search problem over candidate designs. The method assumes as input: (i) a design space \(\mathcal{X}\) containing available subsystem variants; (ii) an executable CPS model \(M(x)\) that can instantiate and simulate a candidate design \(x \in \mathcal{X}\); (iii) functional requirements expressed as signal-based temporal properties over execution traces; and (iv) non-functional design objectives, such as budget, dependability, recoverability, and operational-use criteria. The search targets a high-reward design that satisfies the required temporal behavior and design constraints with a reduced number of executable simulations. In the RL formulation, an action \(a\) corresponds to selecting or updating a candidate design; it is not a runtime control action of the CPS. The selected design is applied to the executable model and simulated under the considered operating conditions. The simulation produces a runtime trace \(\sigma\) containing signals, states, events, fault effects, recovery behavior, timing information, and online monitor outputs. During execution, custom online temporal-property monitors evaluate the relevant signals and produce property-violation indicators. These indicators and the evaluated non-functional terms are combined into the reward value \(r\).

The RL agent receives an observation \(s\) from the executable evaluation environment and selects the next design action according to its current policy. The observation represents the information available to the search process for design selection rather than the complete runtime state of the CPS. For this case study the observation consists of the water and methane levels. The agent action is mapped to the six-dimensional candidate design vector. The selected action changes the candidate design to be evaluated, not the runtime pump-control command. After each evaluation, the transition information \((s,a,r,s')\) is used to update the policy and, for value-based RL, the expected value of design actions. Simulation outcomes are used to update a design-selection policy rather than only for candidate ranking or surrogate-model updates. The transitions associated with previous simulation traces are reused at the level of design choices, and subsequent candidates are not selected as independent samples.

Functional and non-functional requirements are assumed to be available for the target CPS. Functional requirements are translated into signal-based temporal properties over the observable variables exposed by the executable model. This translation depends on the available control, sensing, actuation, and state variables, since only observable or logged signals can be evaluated during execution. The property set is not fixed to the mine-pump case study; for another CPS, different temporal properties can be derived from its functional requirements and evaluated over the corresponding runtime traces. Non-functional requirements are represented through reward terms such as budget, resource consumption, dependability, recoverability, or operational-use criteria.

In this work, the methodology is instantiated in MATLAB/Simulink for the mine-pump CPS. The implementation consists of four components: (i) design-space configuration; (ii) behavioral simulation; (iii) temporal-property evaluation; and (iv) RL-based design search. Figure~\ref{fig:workflow} provides schematics of their interaction through candidate selection, executable evaluation, trace monitoring, reward computation, and feedback-based design search. The following subsections define these components and their case-study instantiation.

\subsection{Design space configuration}
\label{subsec:design-space}

The design space represents the partially known set of CPS designs explored by the search process. Each point in this space represents one candidate design assembled from a set of available design choices. Depending on the target CPS, these choices may correspond to subsystem selection, parameter values, software-to-hardware mappings, communication mechanisms, controller strategies, fault-handling policies, or dependability-related properties. A candidate design can be represented as a configuration vector whose entries correspond to individual design choices. These choices are behaviorally interdependent. A change in one subsystem can modify the conditions under which other subsystems operate and can change the execution dynamics of the CPS. Static configuration values and isolated component scores are insufficient for ranking candidate designs, since structurally similar designs may produce different runtime traces and different temporal-property outcomes. The three configurable subsystems (controller, pump actuator, and methane sensor) can each be instantiated through alternative candidates along one dependability-related and one recoverability-related design dimension. With ten discrete levels per dimension, the number of encoded candidate designs is
\begin{equation}
|\mathcal{X}|
=
(10_{\mathrm{dep}} \times 10_{\mathrm{rec}})^3
=
100^3
=
10^6
=
1{,}000{,}000.
\label{eq:design-space-size}
\end{equation}

The design count reflects only static combinations of design choices. Candidate suitability is determined through executable evaluation, where different fault and recovery trajectories can produce different behavioral outcomes for the same encoded design. For the implemented case study, each candidate design is encoded as a six-dimensional configuration vector:
\[
x =
[x_{C}^{dep}, x_{P}^{dep}, x_{S}^{dep},
 x_{C}^{rec}, x_{P}^{rec}, x_{S}^{rec}],
\]
where \(C\), \(P\), and \(S\) denote the controller, pump actuator, and methane sensor, respectively. The first three entries encode dependability-related choices, and the last three entries encode recoverability-related choices. Each entry takes a discrete value in the range \(1\)--\(10\). The integer levels are abstract design classes used to evaluate the search process. They represent ordered subsystem alternatives with different dependability, recoverability, and cost effects, not direct physical component identifiers. In an industrial study, the same encoding can be replaced by concrete catalog components, architectural options, or parameter ranges. The encoded choices also determine the cost of the assembled design. 
Let
\[
x^{dep} = [x_{C}^{dep}, x_{P}^{dep}, x_{S}^{dep}]
\quad \text{and} \quad
x^{rec} = [x_{C}^{rec}, x_{P}^{rec}, x_{S}^{rec}]
\]
denote the dependability and recoverability parts of the configuration vector. The implementation maps these choices to design costs as
\begin{equation}
C_{k}(x)=\sum_{i \in \{C,P,S\}} \alpha \left(11-x_{i}^{k}\right)^2,
\quad k \in \{\mathrm{dep},\mathrm{rec}\}.
\label{eq:cost-functions}
\end{equation}

The coefficient \(\alpha\) scales the encoded design choices to the budget range used in the experiment. In this case study, a normalized budget limit of \(100\) is used for both dependability-related and recoverability-related choices, with \(\alpha=1.2\). This setting keeps candidate costs within a range suitable for evaluating the search process while still penalizing unfavorable subsystem choices sufficiently. The budget is used as an experimental design constraint, not as a physical procurement value. In other CPS design studies, the same formulation can be connected to real cost, mass, power, or resource budgets, as commonly done in multidisciplinary architecture optimization of systems such as CubeSats~\cite{wijaya2025performance}.

\subsection{Behavioral simulation execution}
\label{subsec:behavioral-simulation}

The behavioral simulation component is the executable counterpart of the design space. A selected configuration is assessed as a running CPS interacting with its operating context, rather than as a static combination of component properties. The simulation includes the configured system, physical process, sensing and actuation paths, controller logic, environmental dynamics, and non-nominal behavior such as faults and recovery. It exposes the execution behavior produced by a selected design.

The main output is a runtime trace \(\sigma\), defined as a time-indexed record of the simulated execution. In the implemented case study, \(\sigma\) includes simulation time, water level, actual and erroneous methane signals, pump activation, controller mode, fault indicators, stuck or progress flags, recovery-related behavior, and pump usage. These signals capture temporal effects that can make a design unsuitable, including fault timing, delayed recovery, subsystem interaction, or late controller response. Dependability and recoverability assumptions are represented through model-based mechanisms~\cite{fabarisov2021model, fabarisov2023remedy}, including FIBlock-based faults in sensing, control, communication, and actuation paths. Since environmental dynamics, fault occurrence, and recovery may be stochastic, repeated executions of the same configuration can produce different traces. During each execution, the online temporal-property monitors observe the relevant runtime signals. The trace records these signals together with the resulting monitor outputs.

\subsection{Temporal-property evaluation}
\label{subsec:temporal-verification}

The temporal-property evaluation component monitors runtime behavior against the functional requirements. Functional requirements are expressed as temporal properties over the signals available in the executable model. The monitored requirements combine LTL-style Boolean temporal structure with STL-style predicates and explicit time bounds over real-valued simulation signals. During simulation, custom online monitors evaluate these properties directly from the runtime trace.

The implementation uses a MATLAB Function block that monitors the runtime trace \(\sigma\) generated by the Simulink model during execution. The block receives the simulation time \(t\), pump activation signal \(p\), actual methane level \(m\), erroneous methane signal \(\hat{m}\), water level \(h\), and stuck flag \(z\). It returns five Boolean violation indicators \(sp_1,\ldots,sp_5\), which provide the temporal-property component of the evaluation later combined with non-functional terms such as cost, recoverability, and operational use. Let \(m_{\mathrm{crit}}\) denote the critical methane threshold, \(m_{\mathrm{safe}}\) the methane level below which pumping is considered safe, \(h_{\mathrm{low}}\) and \(h_{\mathrm{high}}\) the low- and high-water thresholds, and \(\tau\) the reaction-time bound. The monitored properties are:
\[
\begin{aligned}
\varphi_1 &= \mathbf{G}\left(m > m_{\mathrm{crit}} \rightarrow \neg p\right),\\
\varphi_2 &= \mathbf{G}\left((h > h_{\mathrm{high}} \land m < m_{\mathrm{safe}}) \rightarrow \mathbf{F}_{\leq \tau} p\right),\\
\varphi_3 &= \mathbf{G}\left((h < h_{\mathrm{low}} \lor m > m_{\mathrm{safe}}) \rightarrow \mathbf{F}_{\leq 10\tau} \neg p\right),\\
\varphi_4 &= \mathbf{G}\left((h > h_{\mathrm{high}} \land m < m_{\mathrm{safe}}) \rightarrow \neg \mathbf{G}_{\leq \tau}\neg p\right),\\
\varphi_5 &= \mathbf{G}\neg z .
\end{aligned}
\]

Property \(\varphi_1\) specifies methane-related safety: the pump must not remain active under hazardous methane conditions. Property \(\varphi_2\) specifies reactivity by requiring pump activation within the reaction-time bound when water is high and methane is safe. Property \(\varphi_3\) specifies fail-safe shutdown when water becomes low or methane rises above the safe threshold. Property \(\varphi_4\) detects prolonged inactivity under safe high-water conditions, and \(\varphi_5\) detects frozen or unresponsive executions through the stuck flag. For bounded-response properties, the implementation stores trigger times and raises a Boolean violation when the required response does not occur within the corresponding time bound. The resulting indicators \(sp_1,\ldots,sp_5\) are passed to the reward calculation, where repeated violations and sustained compliant behavior are evaluated over the simulation horizon.

\subsection{RL exploration} 
\label{subsec:rl-exploration} 
The RL component performs design-space search. It does not control the mine-pump during operation; the pump remains controlled by the Stateflow controller. Instead, the RL agent selects candidate designs to be evaluated in the executable model. In the RL formulation, the executable CPS model, temporal-property monitors, and reward calculation form the environment. At each search step, the agent observes the current search context \(s_t\), selects a design action \(a_t\), receives reward feedback \(r_t\), and updates its policy from the resulting transition \((s_t,a_t,r_t,s_{t+1})\). The action \(a_t\) represents a design-space decision. In the implemented case study, it is mapped to the six-dimensional candidate design vector \(x_t\) defined in Section~\ref{subsec:design-space}. The action selects or updates dependability-related and recoverability-related design choices for the controller, pump actuator, and methane sensor. It is not a runtime command sent to the pump. The observation \(s_t\) represents the information available to the search process. In this case study, \(s_t\) contains the water and methane levels. These levels are used by the agent as the current search context for selecting the next candidate design; the full CPS execution state remains internal to the Simulink model and is recorded in the runtime trace \(\sigma\). 

A Deep Deterministic Policy Gradient (DDPG) agent is used as the main RL mechanism~\cite{tan2021reinforcement}. DDPG is used here as an actor--critic mechanism for bounded design-action selection, not as a runtime controller. The actor represents the design-selection policy and maps observations to design actions. The critic estimates the expected value of an action under the current observation. The design variables are encoded as bounded integer levels between 1 and 10. DDPG operates on a continuous action representation, and the resulting actions are mapped to the corresponding admissible design levels before simulation. A continuous representation was selected because it preserves proximity relationships between neighboring design alternatives. Candidate designs that differ by one level in a single design dimension remain closer than designs that differ across multiple dimensions. For example, a design encoded as $(10,2,10,10,10,2)$ is more similar to $(9,2,10,10,10,2)$ than to $(1,9,1,1,1,9)$. The critic can therefore generalize value estimates across neighboring regions of the design space rather than treating all design configurations as unrelated discrete choices. The agent stores previously evaluated transitions and reuses them during training. The critic estimates the expected value of design actions and guides policy updates. DDPG combines policy learning with explicit value-function estimation, making it suitable for evaluating the influence of learned value estimates on design-space exploration.

The reward combines functional and non-functional evaluation results. Functional requirements are represented by the Boolean temporal-property violation indicators \(sp_1,\ldots,sp_5\), whose repeated activations are accumulated and severity-weighted over simulation time. The non-functional criteria contribute budget-related terms, recoverability-related terms, sustained-compliance rewards, and an end-of-simulation operational-use bonus. The reward signal is computed as
\begin{equation}
\begin{aligned}
R(t) ={}& R_{dep}(t)-P_{dep}(t)
       + R_{rec}(t)-P_{rec}(t) \\
      & - P_{\varphi}(t)
       + R_{comp}(t)
       + R_{use}(t).
\end{aligned}
\label{eq:reward}
\end{equation}
Here, \(R_{dep}\) and \(P_{dep}\) refer to dependability-related budget fit, \(R_{rec}\) and \(P_{rec}\) to recoverability-related budget fit, \(P_{\varphi}\) to temporal-property violations, \(R_{comp}\) to sustained compliant behavior, and \(R_{use}\) to the final operational-use bonus. The budget terms use the cost mappings introduced in Section~\ref{subsec:design-space} and penalize both over-budget designs and designs below the minimum design threshold. The temporal-property penalty uses the violation indicators from Section~\ref{subsec:temporal-verification}:
\[
P_{\varphi}(t)=
\sum_{i=1}^{5} w_i sp_i(t),
\quad
w=[5.0,\;1.5,\;4.2,\;1.0,\;1.8].
\]
The weights reflect the relative severity assigned to the monitored violations in this case study and can be adapted to the priorities of another CPS design problem. Cooldown intervals control the repeated accumulation of property violations and sustained-compliance rewards over simulation time. The reward combines Boolean monitor outputs with time-accumulated temporal and behavioral evidence over the simulation horizon.

\section{Experiments and discussion}
\label{sec:experiments}
The experiments evaluate the hypothesis stated in Section~\ref{sec:research-gap}: experience accumulated across previous simulation traces can be reused to reduce the simulation effort required for design-space exploration. The evaluation compares RL-guided design search with GA-based and BO-based search under the same executable setup and then uses ablation variants to identify which RL mechanisms explain the observed performance differences. A brute-force baseline is not used since the implemented design space contains \(10^6\) encoded candidate designs before repeated stochastic executions are considered. The evaluation focuses on representative population-based, surrogate-guided, and RL-guided search strategies, together with ablation variants that assess the role of value-based feedback and reuse of previous execution traces.

\subsection{Experimental setup}
\label{subsec:experimental-setup}
The experiments use the mine-pump CPS introduced in Section~\ref{sec:motivating-example} and the six-dimensional design encoding defined in Section~\ref{subsec:design-space}. Each evaluated design is executed in the same stochastic Simulink model and assessed using the same temporal-property indicators and reward calculation from Section~\ref{subsec:rl-exploration}.

The evaluation consists of a comparative study and an ablation study. The comparative study evaluates RL-guided design search against GA-based and BO-based search. The ablation study investigates which RL mechanisms contribute to the observed performance differences.  All methods are evaluated under a common executable model, design encoding, and reward formulation. GA-based search is used as a representative population-based baseline, whereas BO is used as a representative surrogate-guided baseline. Both methods operate on the same encoded design space and use the same reward calculation as the RL-guided approach. The comparison evaluates simulation effort and achieved reward quality across the considered search mechanisms. Let \(\bar{R}_n\) and \(s_n\) denote the sample mean and sample standard deviation of the final best reward over \(n\) independent optimization runs. The relative half-width of the normal-approximation \(95\%\) confidence interval is

\[
\delta_n=\frac{1.96\,s_n}{\sqrt{n}\,|\bar{R}_n|}.
\]
BO and GA were each executed for at least \(20\) runs and stopped when \(\delta_n \leq 0.10\) (subject to a maximum of \(50\) runs).

\begin{figure}[t]
    \centering
    \begin{subfigure}[t]{0.48\columnwidth}
        \centering
        \includegraphics[width=\linewidth]{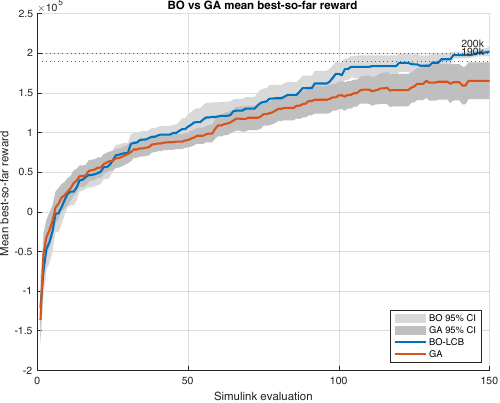}
        \caption{Best-so-far reward progression.}
        \label{fig:bo-ga-progress}
    \end{subfigure}
    \hfill
    \begin{subfigure}[t]{0.48\columnwidth}
        \centering
        \includegraphics[width=\linewidth]{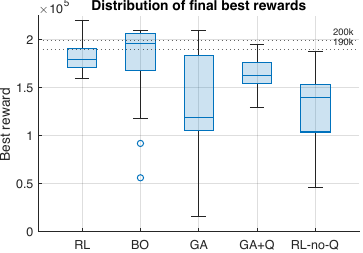}
        \caption{Distribution of final best rewards.}
        \label{fig:bo-ga-boxplot}
    \end{subfigure}

    \vspace{0.5em}

    \begin{subfigure}[t]{\columnwidth}
        \centering
        \includegraphics[width=0.85\linewidth]{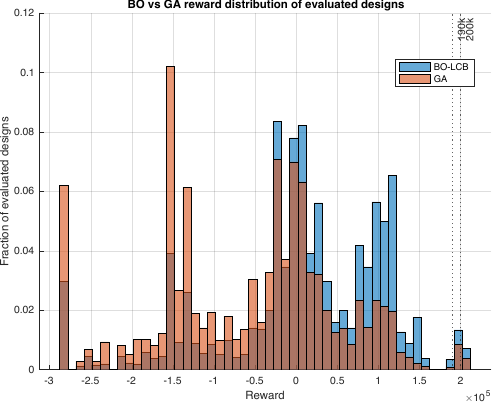}
        \caption{Reward distribution of evaluated designs.}
        \label{fig:bo-ga-reward-distribution}
    \end{subfigure}

\caption{Comparative and ablation results for simulation-based CPS design search.
(a) Search efficiency measured by the best reward obtained as a function of the evaluation budget for BO and GA.
(b) Distribution of final best rewards across the evaluated search variants.
(c) Distribution of rewards observed during search for BO and GA.}
    \label{fig:bo-ga-comparison}
\end{figure}

The main comparison considers three design exploration methods:
\begin{enumerate}
    \item RL-guided design search using the DDPG agent described in Section~\ref{subsec:rl-exploration};
    \item GA-based search using the same encoded design space and reward calculation;
    \item BO-based search using the same encoded design space and reward calculation.
\end{enumerate}

The ablation study investigates which RL mechanisms contribute to the observed performance differences:
\begin{enumerate}
    \item actor-only RL without a Q-function, used to assess the role of value-based feedback;
    \item Q-informed GA, where previous evaluations are used to estimate candidate quality before simulation.
\end{enumerate}

The evaluation considers both simulation effort and achieved reward quality. Simulation effort is measured as the number of executable Simulink evaluations required before a method identifies a high-reward design. Reward quality is assessed through the best design reward obtained during the search process.

\subsection{Comparative evaluation}
\label{subsec:rl-ga}
The comparative study evaluates RL-guided design search against GA-based and BO-based search under the common executable evaluation setup defined in Section~\ref{subsec:experimental-setup}. All methods use the same encoded design space, temporal-property indicators, and reward calculation.

\textbf{Observation 1: BO achieves higher reward quality and consistency than GA.}
Figure~\ref{fig:bo-ga-progress} compares the best-so-far reward progression of BO and GA as a function of the evaluation budget. Both methods improve rapidly during the initial evaluations. BO exhibits continued improvement over a larger portion of the evaluation budget, whereas GA shows slower reward growth.
Figure~\ref{fig:bo-ga-boxplot} summarizes the distribution of final best rewards obtained across independent optimization runs. BO achieved a mean best reward of 176,765 and a median best reward of 196,504, compared with 130,993 and 118,931 for GA, respectively. The best solutions discovered by both methods were similar, reaching approximately 209,000 reward units. The effect size between the two methods was large ($d=0.98$). Both the rank-sum test ($p=4.94 \times 10^{-4}$) and the two-sample $t$-test ($p=1.55 \times 10^{-4}$) indicated statistically significant differences. 

Figure~\ref{fig:bo-ga-reward-distribution} presents the distribution of rewards observed during search. BO evaluated a larger proportion of designs in high-reward regions, whereas GA evaluated a larger proportion of designs in lower-reward regions. BO achieved rewards above 190,000 in 68\% of runs and rewards above 200,000 in 36\% of runs. The corresponding success rates for GA were 24\% and 10\%, respectively. BO satisfied the stopping criterion after \(25\) runs with \(\delta_{25}=0.09756\). GA reached the \(50\)-run cap with \(\delta_{50}=0.1019\). \\

\textbf{Observation 2: RL reaches high-reward designs with fewer executable evaluations.}
The RL-guided workflow identified the highest-reward design found after 26 episodes and converged within 32 episodes. Under the simulation accounting used in this experiment, this corresponds to 130 executable simulations.
The reduction does not originate from differences in the cost of individual evaluations. All methods execute the same Simulink model and use the same temporal-property indicators and reward calculation.

\subsection{Ablation and mechanism analysis}
\label{subsec:ablation}

The empirical evaluation showed that the RL-guided workflow exhibits a different optimization behavior than the GA-based and BO-based search methods. The ablation study investigates which RL mechanisms contribute to the observed differences.
The ablation study evaluates two components of the RL workflow: reuse of previous simulation outcomes and value-based feedback for design-action selection. In DDPG, previous interactions are stored as tuples $(s_t,a_t,r_t,s_{t+1})$ and reused during training. The critic provides value-based feedback by estimating the expected cumulative reward associated with a design action. The actor uses this estimate to update the design-selection policy.

\textbf{Observation 3: Learned value estimates improve design-space exploration.}
Figure~\ref{fig:bo-ga-boxplot} includes the full RL workflow, the actor-only RL variant, the standard GA, and the Q-informed GA variant. Removing the Q-function leads to slower and less stable learning. Although the actor-only variant continues to improve over time, it requires more episodes to reach high-reward designs and exhibits less consistent performance.
The Q-informed variant uses previously evaluated candidates to estimate the expected reward of new candidates before simulation. This modification improves search performance compared with the standard reward-driven GA.
Learned value estimates derived from previous evaluations improve search effectiveness. Similar principles are also present in BO, where a surrogate model is updated using previous observations, and in RL, where replay and critic updates reuse prior experience. The methods differ in how this information is incorporated into the search process.

\textbf{Observation 4: The combination of replay, value estimation, and policy learning contributes to the observed performance improvements.}
Neither the actor-only RL variant nor the Q-informed GA variant reproduces the behavior of the full RL workflow. The actor-only variant lacks critic-based value estimation, while the Q-informed GA variant uses value estimates only to guide candidate selection within a population-based search process. In contrast, the full RL workflow combines replay-based experience reuse, critic-guided value estimation, and policy optimization within a single learning framework.
The actor-only RL variant isolates policy learning without critic guidance. The Q-informed GA variant isolates value estimation within a population-based search process. The full RL workflow combines both mechanisms together with replay-based experience reuse.

\section{Conclusion and future work}
\label{sec:conclusion}
This paper presented an RL-guided workflow for temporal-property-driven design-space exploration of configurable CPS. Candidate designs are assembled from subsystem alternatives and evaluated through executable simulation. During each evaluation, online temporal-property monitors produce functional-property violation indicators from the runtime trace. These indicators and the evaluated non-functional terms determine the reward. The model-based workflow comprises design-space configuration, behavioral simulation, temporal-property monitoring, reward calculation, and candidate-design selection. The executable mine-pump CPS model is made publicly available. The evaluation compared RL-guided search with GA and BO under the same executable model, temporal-property indicators, and reward formulation. BO achieved higher final reward quality and rates of high-reward solutions than GA. RL reached the highest-reward design identified in the experiments with fewer executable simulations than the baselines. The ablation study showed that value estimation, replay-based experience reuse, and policy learning contribute to search performance. Because RL training is substantially more computationally expensive, the same repeated-run statistical protocol was not applied to all three methods: RL was evaluated through the available training campaign, whereas GA and BO were assessed over repeated runs. This limits statistical comparability and motivates a broader evaluation with systematic evidence collection.

BO, Q-informed GA, and RL reuse previous executable evaluations through different mechanisms. Two complementary lines of work are currently pursued. The first structures and interprets evidence from successful and unsuccessful candidates, including design vectors, simulation traces, monitor outputs, temporal-property outcomes, and rewards, to synthesize new design candidates. The second automates candidate execution, evidence collection, and reuse of new evidence across repeated design-exploration cycles. This extends the present approach beyond a single DSE campaign. The mine-pump case study will be extended to more complex CPS behavior. Scalability will be evaluated with larger design spaces and additional temporal properties, and generalizability beyond MATLAB/Simulink will be assessed using further executable CPS case studies, including Python.

\noindent\textbf{Acknowledgment:} This work is supported by the Luxembourg National Research Funds (FNR) through the project grant C23/IS/18177547/VARIANCE.

\bibliographystyle{IEEEtran}
\bibliography{main}
\end{document}